# Does hybrid density functional theory predict a non-magnetic ground state for δ–Pu?

Raymond Atta-Fynn and Asok K. Ray*

*Physics Department, University of Texas at Arlington, Arlington, Texas 76019*

*akr@uta.edu.

# Abstract

Hybrid density functionals, which replaces a fraction of density functional theory (DFT) exchange with exact Hartree-Fock (HF) exchange, have been used to study the structural, magnetic, and electronic properties of δ-Pu. The fractions of exact Hartree-Fock exchange used were 25%, 40%, and 55%. Compared to the pure PBE functional, the lattice constants expanded with respect to the experimental value when the PBE-HF hybrid functionals were applied. For pure PBE and hybrids functionals with HF exchanges amounts of 25% and 40%, the ground state structure was anti-ferromagnetic, while for 55% HF contribution the ground state was non-magnetic. The *5f* electrons tend to exhibit slight delocalization or itinerancy for the pure PBE functional and well-defined localization for the hybrid functionals, with the degree of *5f* electron localization increasing with the amount of HF exchange. Overall, the performance of the hybrid density functionals do not seem superior to pure density functionals for δ – Pu.

## 1. Introduction

The actinides- a group of fourteen elements with atomic numbers 90 to 103 that follow Actinium in the periodic table- are characterized by the gradual filling of the valence 5$f$ shells and have been the subject of intense experimental and theoretical research for various scientific, technological, and societal reasons [1–3]. Among the actinides, plutonium (Pu) occupies a crucial position as elements to the left (thorium to neptunium) have itinerant 5$f$ electrons and the elements to the right (americium and beyond) have localized 5$f$ electrons. Thus Pu is at the boundary where 5$f$ electrons delocalization to localization transition in the actinides occurs. Pu is a chemically and physically complex metal with six known different crystal structures between room temperature and the anomalously low melting point of 913 K under atmospheric conditions. $\delta$-Pu, in particular, is most technologically important because it is ductile and malleable (the ground state $\delta$-Pu is hard and brittle) and can be stabilized at room temperature when alloyed with a few atomic percent of some impurities such as Ga and Al. However, $\delta$-Pu, though being the most studied phase, is probably the least understood phase due to the nature of the 5$f$ electrons, which are believed to have intermediate properties between itinerant and localized behavior.

One of the very puzzling questions regarding Pu relates to the question of magnetism. Experimental studies have shown that there is no evidence of ordered or disordered moments at low temperatures, in either the $\alpha$ or $\delta$ phase [4, 5]. Using photoelectron spectroscopy (PES), Arko *et al.* [6] probed the electronic structure of $\alpha$-Pu and $\delta$-Pu and observed that the 5$f$ states in both phases display temperature independent, narrow bands near the Fermi level with the δ-Pu band being narrower than that of α-Pu. Using ultraviolet photoelectron spectroscopy, Gouder *et al.* [7] obtained a valence band spectra of δ-Pu similar to the one in Ref. [6]. By measuring the branching ratio of core-valence transitions in electron-energy-loss-spectra (EELS), Van der Laan *et al.* [8] and Moore *et al.* [9] showed that the valence is at or near $5f^5$ and the coupling mechanism is exactly intermediate, being very close to the *jj* limit.

Theoretically, there is no unified picture for δ-Pu. Density functional theory (DFT) calculations of δ-Pu within the local density approximation (LDA) or the generalized gradient approximation (GGA) [10–16] produce a magnetic ground state (usually anti-ferromagnetic) with lattice constant close to experiment and sizable magnetic moments. On the other hand, non-magnetic DFT solutions underestimate the experimental lattice constant by as much as 30% and predict a bulk modulus which is four times larger than experimental value. Furthermore, these calculations do not capture the strong localization characteristics of the 5*f* electrons in the vicinity of the Fermi level at the AFM ground state as predicted by experiment [6, 7]. The LDA+U approach predicts a strongly magnetic $f^5$ configuration [17, 18] or a nonmagnetic $f^6$ configuration [19, 20] for the localized *f* shell depending on the values of certain parameters. For the nonmagnetic $f^6$ δ-Pu ground state [19, 20], the equilibrium atomic volume was in good agreement with experiment but the 5*f* computed density of states (DOS) failed to match experimental PES data [6, 7]. Using LDA+DMFT, Savrasov *et al.* [21] matched the experimental PES of δ-Pu. DMFT results, just like that of LDA+U, depend on U and for material like Pu with six phases, there is not a single U that can describe all phases. Effectively, U becomes an adjustable parameter that must be varied to yield the desired result. Shim *et al.* [22] observed valence fluctuations between $f^4$, $f^5$, and $f^6$ for α-Pu and δ-Pu, with the average 5$f$ occupancy being 5.2; Zhu *et al.* [23] studied the sensitivity of the δ-Pu 5$f$ occupancy to its spectral properties and concluded that the 5$f^5$ configuration gives the best match with experimental PES data.

In view of the above discussions and some recent successes of applying hybrid density functional theory to actinide compounds [24], the purpose of this work is to explore this theory for elemental actinides like δ-Pu and specifically investigate the dual questions pertaining to magnetism and localization. Using the PBE0 hybrid functional [25] (also known as PBE1PBE), which combines 25% exact Hartree-Fock (HF) exchange with 75% of Perdew-Burke-Ernzerhof (PBE) exchange, and uses the PBE correlation functional as formulated within GGA [26], Kudin *et al.* [24] obtained the correct experimental anti-ferromagnetic insulating ground state of $UO_2$, whereas pure DFT predicted a ferromagnetic metallic ground state. Also, for $PuO_2$, a non-zero band gap, a direct consequence of 5$f$ electron localization, was observed in PBE0 hybrid DFT [25] calculations by Prodan *et al.* whereas conventional DFT produced a zero band gap [24].

**2. Computational method and discussions of results**

All computations were carried out within the full-potential linear augmented plane wave plus local (FP-LAPW+lo) basis method in the Wien2k program [27]. The unit cell is divided into non-overlapping muffin-tin spheres and an interstitial region. This method makes no shape approximation to the charge density and potential in the sense that non-spherical matrix elements are added to the charge density and potential inside the muffin-tin by expanding them in terms of lattice harmonics and in the interstitial region the expansion is carried as a Fourier series. Inside the muffin-tin sphere of radius $R_{MT}$, the wave functions are expanded using radial functions times spherical harmonics with angular momentum up to $l$=10 and in the interstitial region the expansion is carried out using planes waves. The muffin-tin radius was set at $R_{MT}$(Pu)=2.7 a.u. and the plane wave convergence parameter was set to $R_{MT}K_{MAX}$=9. Additional local orbitals (LO) were added to the 6$s$ and 6$p$ semi-core states to improve their description. For more technical details about the Wien2k code, readers should refer to Ref. [27] and the references therein.

For $δ$-Pu, the fcc unit cell with a single atom is sufficient to model the non-magnetic (NM) and ferromagnetic (FM) configurations. To model the anti-ferromagnetic (AFM) configuration, a 1 × 1 × 2 supercell was used (unit cell was doubled along one coordinate direction), with the two neighboring atoms having anti-parallel spins. This changed the fcc unit cell to a tetragonal unit cell. For the sake of consistency, the 2-atom tetragonal unit cell was used for all calculations. Brillouin zone integration was performed using about a total of 1000 k-points. The self-consistent field iterations were terminated when the total energy and charge density converged to $10^{-5}$ Ry and $10^{-3}$, respectively or better. All computations included spin-orbit coupling and orbital polarization [28,29]. The spin quantization axis was along the [001] direction.

Recently, Novák *et al.* [30] proposed an improvement of the description of strongly correlated electrons by subtracting the LDA or GGA exchange-energy functional corresponding to the subspace of the states of the correlated electrons and then added the HF expression instead. This method, called "exact exchange for correlated electrons," was implemented within the FP-LAPW+lo method and it was successfully applied to several 3*d* and 4*f* systems. Details of the implementation have been discussed in a recent article by Tran *et al.* [31]. Within this implementation, the hybrid functional is applied but only to a selected set of electrons inside the muffin tin, namely the ones that are poorly treated by LDA and GGA, which in this work will be the Pu 5*f* electrons. Specifically, the hybrid exchange-correlation energy functional $E_{XC}^{PBE+\alpha HF}$, used in this work, has the form

$$E_{XC}^{PBE+\alpha HF}[\rho] = E_{XC}^{PBE}[\rho] + \alpha\left(E_X^{HF}[\Psi_{5f}] - E_X^{PBE}[\rho_{5f}]\right), \qquad (1)$$

where $E_{XC}^{PBE}$ is the PBE exchange-correlation functional [26], $E_X^{HF}$ is the HF exchange functional, $E_X^{PBE}$ is the PBE exchange functional, $\Psi_{5f}$ and $\rho_{5f}$ are the wave function and the corresponding electron density of the 5*f* electrons respectively, $\rho$ is the total electron density, and the parameter $\alpha$ denotes the fraction of HF exchange replacing the PBE exchange but only for the selected atoms. For this work we used $\alpha$=25%, 40%, and 55%. To ensure numerical accuracy, the numerical mesh inside the muffin tin sphere, which was used for the computation of the exact exchange, was made 2.5 times denser. Test calculations for α=55% indicated that the largest change in the equilibrium volume and the relative energy differences between the different magnetic structures from the results reported here was less than 0.4 %. This implies that the default numerical mesh is sufficiently accurate.

In Table 1, we report the total energy, equilibrium lattice constants, bulk modulus, and spin, orbital, and total magnetic moments of $\delta$-Pu for the pure PBE density functional and hybrid density functionals. Fig. 1 indicates that the ground state structures for the PBE, PBE+25% HF, and PBE+40% HF functionals are AFM, all of which disagree with experiments. On the other hand, the PBE+55% HF functional yields an NM ground state, which agrees with experimental results. For the pure PBE functional, the energy differences, particularly the AFM-NM difference (36.47 mRy/atom), are quite significant, and the magnitudes are in agreement with previous work [15,29]. With the inclusion of 25% HF exchange, the energy of the AFM-NM and FM-AFM energy differences increase and decrease to 63.33 mRy/atom and 5.24 mRy/atom, respectively. For the PBE+40% HF functional, the AFM-NM energy difference decreases significantly to7.35 mRy/atom. Clearly for 55% HF contribution for which the ground state is NM, the NM-AFM and NM-FM energy differences of 82.28 mRy/atom and 86.33 mRy/atom are the largest compared to each theoretical level. The sudden stabilization of the NM configuration from 25% HF to 40% HF contributions and the large NM-FM/AFM energy differences for 55% HF contribution suggests that for higher fraction of HF exchange, the NM ground state may still be realized. We tried to verify this by performing calculations using a hybrid functional with 70% HF exchange but failed to achieve convergence.

For the PBE and PBE+25% functionals, the magnitude of the spin moments slightly exceed that of the orbital moments and overall, a large cancellation of the spin and orbital moments occurs, leading to a small total magnetic moment. For the pure PBE functional, the two magnetic cells have net magnetic moments 0.88 $\mu$B (FM) and 0.96 $\mu$B (AFM). Of particular interest is the AFM cell at the PBE+25% level with nearly vanishing total moment of 0.07 $\mu$B. It should be noted that the orbital moment is computed only inside the muffin sphere (no interstitial contribution), which comprises 40.6% of the total volume. We expect the total moment to be even smaller or completely vanish for large muffin tins. Thus for the PBE+25% HF functional, we conclude that $\delta$-Pu is a magnetic metal but with no net moments present. For 40% and 55% HF contributions, |ML| > |MS| and the resulting total moments are still fairly small but significantly larger than the PBE and PBE+25% HF functionals. In particular, the formation of magnetic moments for the FM and AFM cells at 55% HF contributions tends to destabilize the structure considering the energy differences.

In Fig. 1, we depict total energy versus atomic volume plots for each magnetic configuration per density functional. The data was fitted using the Birch-Murnaghan equation of state to obtain the equilibrium atomic volume and bulk modulus [33]. For the PBE functional the equilibrium lattice constant is underestimated for the NM cell, while the FM and AFM lattice constants are in satisfactory agreement with experimental data. With 25% HF contribution, the lattice constant of the NM configuration expands by 0.52 a.u. compared to the PBE functional and deviates from experiment by 1.3%. For the remaining cases, the lattice constants expand considerably compared to experimental data, with percent differences ranging from 4.1%-7.6%. In terms of atomic volume, this range translates to 12.8%-24.7% volume expansion compared to the experimental volume. We note that increasing 5$f$ electron localization in the actinides, which implies increasing the non-participation of the 5$f$ electrons in bonding, leads to increasing volume expansion. Only four bulk moduli, namely FM and AFM for the PBE and PBE+25% HF functionals show a good or fair agreement with experiment. All others show large deviations from experiment. Bulk moduli are known to notoriously deviate from experimental data, and usually experimental data indicate large uncertainties.

In Table 1, the partial $f$-electron population is reported. We wish to point out that the charge contribution to the interstitial region cannot be resolved per each angular momentum channel and that the partial charges inside the sphere depend on the radius. Clearly for the pure DFT-PBE case, the $f$ electron count is close to 5, while for the 25% HF hybrid functional the $f$ electron count reduces to 4.7 for the AFM and FM structures and 4.4 for the NM structure. For the 40% HF and 55% HF, the Pu atom has basically an $f^4$ valence. Clearly the decrease in the $f$ electron population inside the muffin tin sphere with respect to increasing HF fraction is a direct consequence of the anomalously large $f$ electron localization and the increasing total energy. For the 55% HF fraction, it is plausible to claim that the on-site exchange interaction and the $f^4$ valence are the agents that drive the NM state to stabilization. A NM 5$f^4$ $\delta$-Pu valence contradicts experimental value close to $5f^5$ [8, 9]. As stated earlier, LDA+U yield a NM solution for δ-Pu with a $5f^6$ valence [19, 20], while LDA+DMFT produces a NM solution with $5f^{5.2}$ valence [21–23].

We also computed the 5f electronic density of states (DOS) for each functional and they are presented in Figs. 2-5. In Fig. 2, where we have plotted the DOS for the PBE functional, we observe that the NM DOS exhibit a sharp peak just below the Fermi level and a sudden drop at the Fermi, implying a signature of some 5f electron localization. For the AFM and FM DOS, the bands show relatively broad characteristics, signifying itinerant character. Figs. 3-5 clearly reveal that the onset of the HF functional results in well-localized 5f states. Looking at the DOS for the PBE+25% HF functional presented in Fig. 3, we see a withdrawal of the 5f states from the Fermi level with a well-defined sharp peaks in the valence region, leading to a reduction in the DOS at the Fermi level compared to Fig. 2, coupled with a 2.5 eV splitting between the $5f_{5/2}$ and $5f_{7/2}$ bands at the AFM ground state. For the DOS of PBE+40% HF in Fig. 4, the 5f states are pushed down further below the Fermi level with well-localized with states centered at 3.7 eV and the splitting between the $5f_{5/2}$ and $5f_{7/2}$ bands increase to 4.2 eV and a further reduction in the DOS at the Fermi level. The localization increases slightly from PBE+40% HF to PBE+55% HF as can be seen in Fig. 5, with the NM ground state showing the most 5f electron localization, with the $5f_{5/2}$ and $5f_{7/2}$ bands splitting by 8.5 eV and virtually no DOS at the Fermi level. Obviously, all the DOS presented in Figs. 2-5 fail to match experimental PES data [6, 7], particularly Figs. 3-5, where there is a large withdrawal of the 5f states from the immediate vicinity of the Fermi level. It is possible that the splitting of the occupied 5f states in the DOS curves for the 40% and 55% HF hybrid functionals is due to crystal field interaction. However, the DOS curves indicate that spin-orbit interaction is much stronger than that of the crystal field. The significant 5f electron localization observed in Figs. 3-5, which implies non-participation of the 5f electrons in bonding, provides the answer to the question as to why the lattice constant expands when the hybrid functionals are applied. To explain the large withdrawal of the 5f bands from the Fermi level, we first note that the HF method is known to perform poorly when applied to metals (e.g. vanishing DOS at the Fermi energy) [34]. Another artifact of the HF method is the overestimation of band gaps. Thus the anomalously strong 5f electron localization and subsequently, the splitting between the occupied and unoccupied bands, as well as the reduction of the DOS at the Fermi level are manifestations of the artifacts of HF theory.

The HF artifacts stems from the unscreened exchange interaction. One possibility is certainly to use the "screened" hybrid density functional due to Heyd *et al.* (HSE) [35], which employs a screened, short-range HF exchange instead of the full exact exchange developed with the aim of avoiding the problems mentioned above. For non-actinide metals, HSE improves upon pure DFT results, but the effect is less dramatic compared to band gap calculations of semiconductors where the use of HSE produces errors over five times smaller than pure DFT results. Since the HSE functional is not yet implemented in Wien2k, we are unable to access its impact on δ-Pu or any other actinides at this point. We intend to pursue this in the future. In a related context, it is worth noting that Paier *et al.* [36] have observed that another hybrid functional B3LYP performs worse than PBE0 or HSE functional for extended metallic systems.

In summary, we have used a hybrid functional which combines exact HF exchange with the PBE exchange-correlation functional to study the structural, magnetic, and electronic properties of *δ*-Pu for 25%, 40%, and 55% exact HF exchange contributions. For 55% HF contribution, the ground state is non-magnetic, otherwise it is AFM. Furthermore a large cancellation of the spin moments by the orbital moments was observed in the magnetic cells. Save some few cases, the bulk modulus show significant deviation from experiments. The hybrid functionals leads to the expansion of the lattice constants due to significant 5*f* electron localization. We conclude that though the hybrid functionals work well in the actinide oxides, they do not show marked improvements in predicting the properties of *δ*-Pu when compared to pure DFT. We expect this conclusion to hold true for the other phases of Pu.

## Acknowledgments

This work is supported by the Chemical Sciences, Geosciences and Biosciences Division, Office of Basic Energy Sciences, Office of Science, U. S. Department of Energy (Grant No. DE-FG02-03ER15409) and the Welch Foundation, Houston, Texas (Grant No. Y-1525). This research also used resources of the National Energy Research Scientific Computing Center, Office of Science, U.S. Department of Energy (Contract No. DE-AC02-05CH11231) and the Texas Advanced Computing Center (TACC).

Table I: Equilibrium properties of δ-Pu for different density functionals: PBE0 denotes PBE+25% HF, PBE2 denotes PBE+40% HF, PBE3 denotes PBE+55% HF, E is the total energy, $M_S$ is the spin magnetic moment, $M_L$ is the orbital magnetic moment, $M_J = M_S + M_L$, $a$ is the lattice constant, $\Delta a$ is the percent difference in the lattice constant, B is the bulk modulus, and $n_{5f}$ is the partial f-electron population (computed inside the muffin tin sphere). For each density functional, the total energy of the lowest energy structure is set to zero.

|  | Magnetic Configuration | E (mRy/atom) | $M_S$ ($\mu_B$) | $M_L$ ($\mu_B$) | $M_J$ ($\mu_B$) | $a$ (a.u.) | $\Delta a$ (%) | B (GPa) | $n_{5f}$ |
|---|---|---|---|---|---|---|---|---|---|
| PBE | NM | 36.47 | 0 | 0 | 0 | 8.35 | -4.7 | 62.6 | 4.9 |
| PBE | FM | 11.82 | 4.24 | -3.36 | 0.88 | 8.84 | 0.9 | 28.3 | 4.8 |
| PBE | AFM | 0 | 4.15 | -3.19 | 0.96 | 8.75 | 0.1 | 54.0 | 4.8 |
| PBE0 | NM | 63.33 | 0 | 0 | 0 | 8.87 | 1.3 | 69.6 | 4.4 |
| PBE0 | FM | 5.24 | 5.22 | -4.62 | 0.60 | 9.43 | 7.6 | 40.8 | 4.7 |
| PBE0 | AFM | 0 | 4.67 | -4.60 | 0.07 | 9.33 | 6.5 | 43.0 | 4.7 |
| PBE2 | NM | 7.35 | 0 | 0 | 0 | 9.12 | 4.1 | 79.8 | 4.0 |
| PBE2 | FM | 3.87 | 3.82 | -5.35 | -1.53 | 9.12 | 4.1 | 85.2 | 4.1 |
| PBE2 | AFM | 0 | 3.92 | -5.50 | -1.58 | 9.15 | 4.5 | 76.3 | 4.1 |
| PBE3 | NM | 0 | 0 | 0 | 0 | 9.28 | 5.9 | 77.5 | 3.9 |
| PBE3 | FM | 86.33 | 4.04 | -5.68 | -1.64 | 9.24 | 5.5 | 76.4 | 4.0 |
| PBE3 | AFM | 82.28 | 4.04 | -5.70 | -1.66 | 9.24 | 5.5 | 72.3 | 4.0 |
| Experiment | NM[a] |  | N/A | N/A | 0 | 8.76[b] |  | 29-35[c] |  |

[a]Ref. [4,5]
[b]Ref. [1]
[c]Ref. [32]

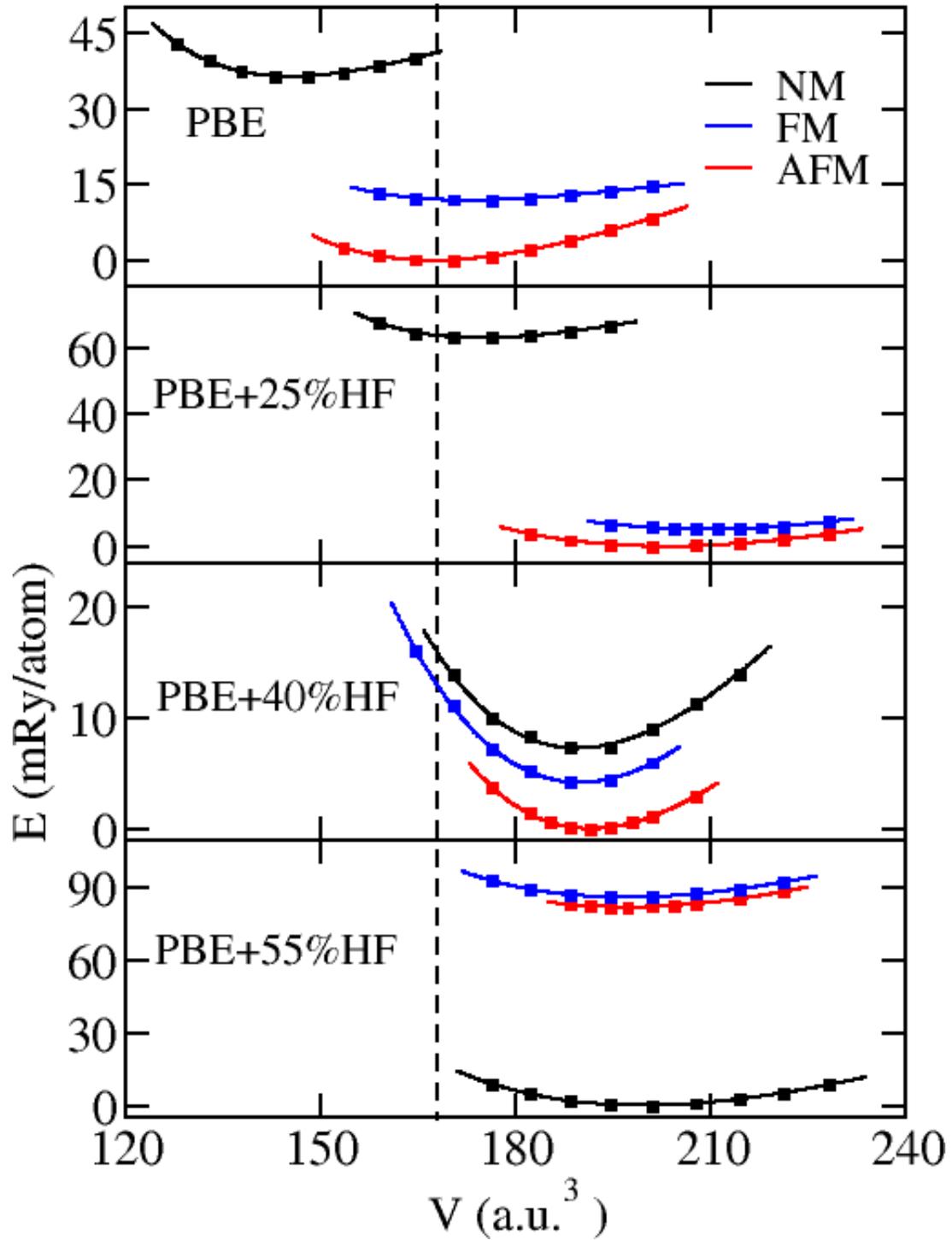

Fig. 1: Total energy (E) as a function of atomic volume (V) for δ-Pu for different functionals. The dashed vertical line denotes the δ-Pu experimental volume of 168.06 a.u.$^3$.

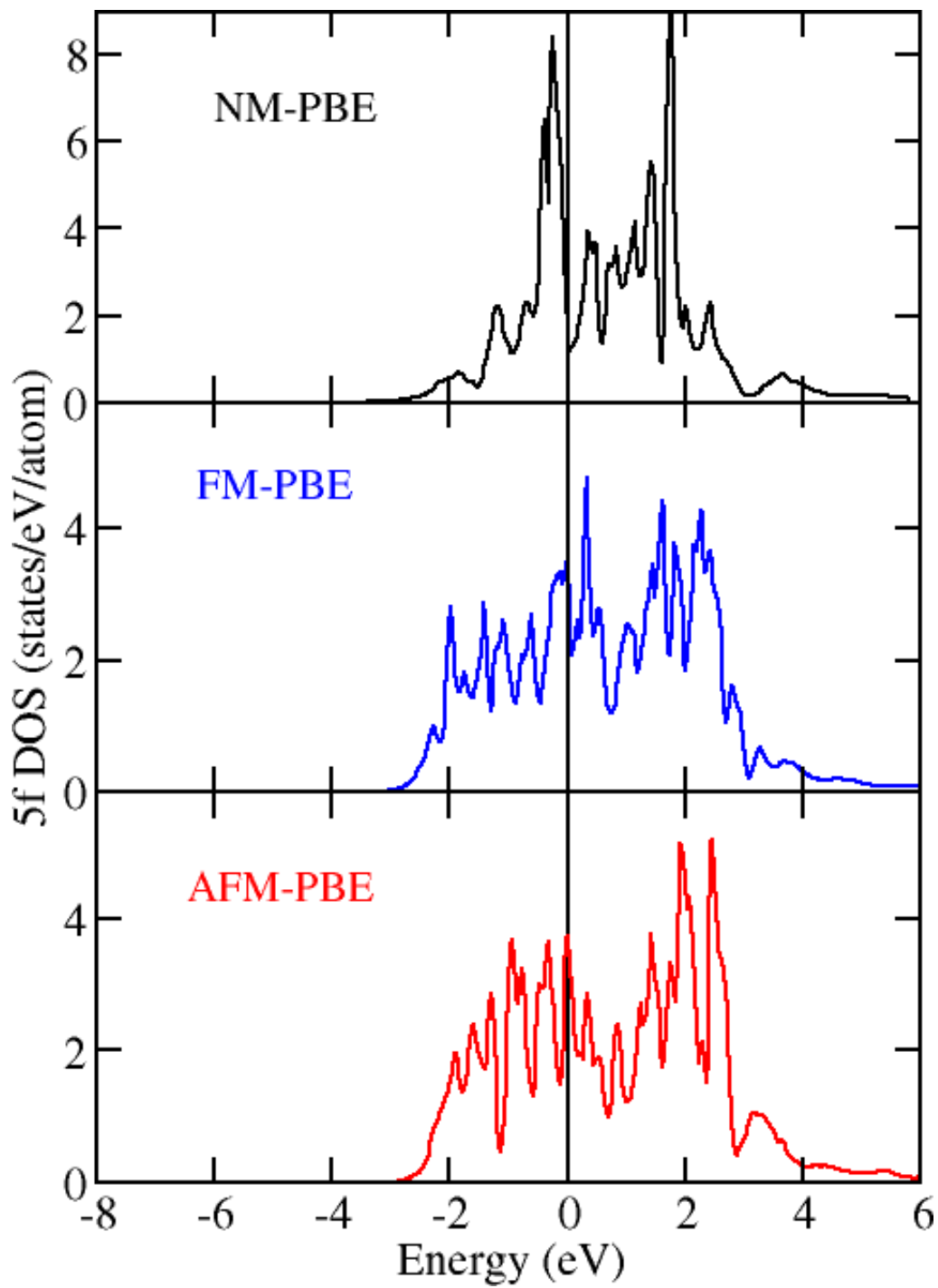

Fig. 2: 5f partial density of states for δ-Pu for the non-magnetic (NM), ferromagnetic (FM), and anti-ferromagnetic (AFM) computed using the PBE functional. Solid vertical line through zero energy is the Fermi level.

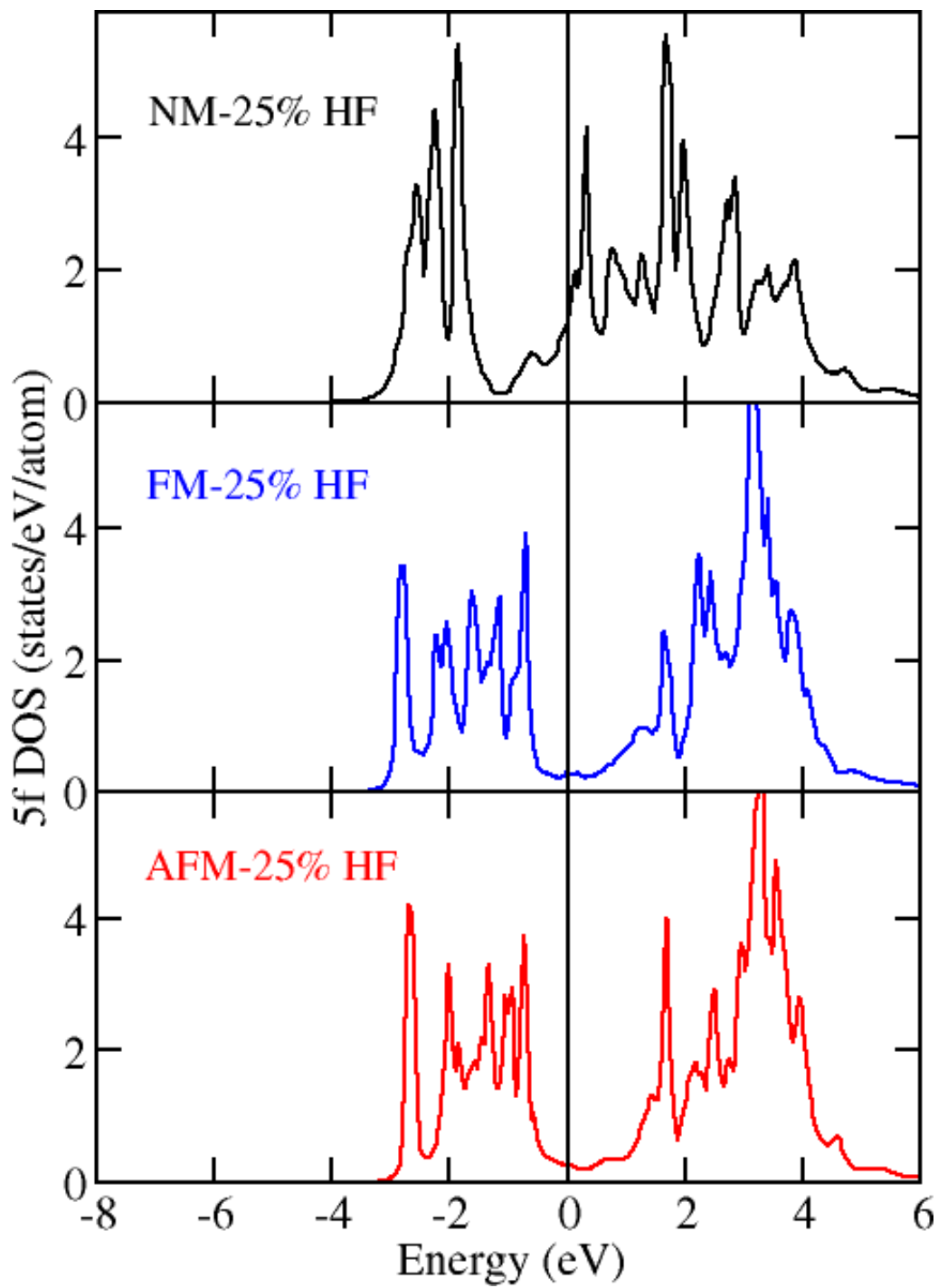

Fig. 3: *5f* partial density of states for δ-Pu for the non-magnetic (NM), ferromagnetic (FM), and anti-ferromagnetic (AFM) computed using the PBE0 functional.

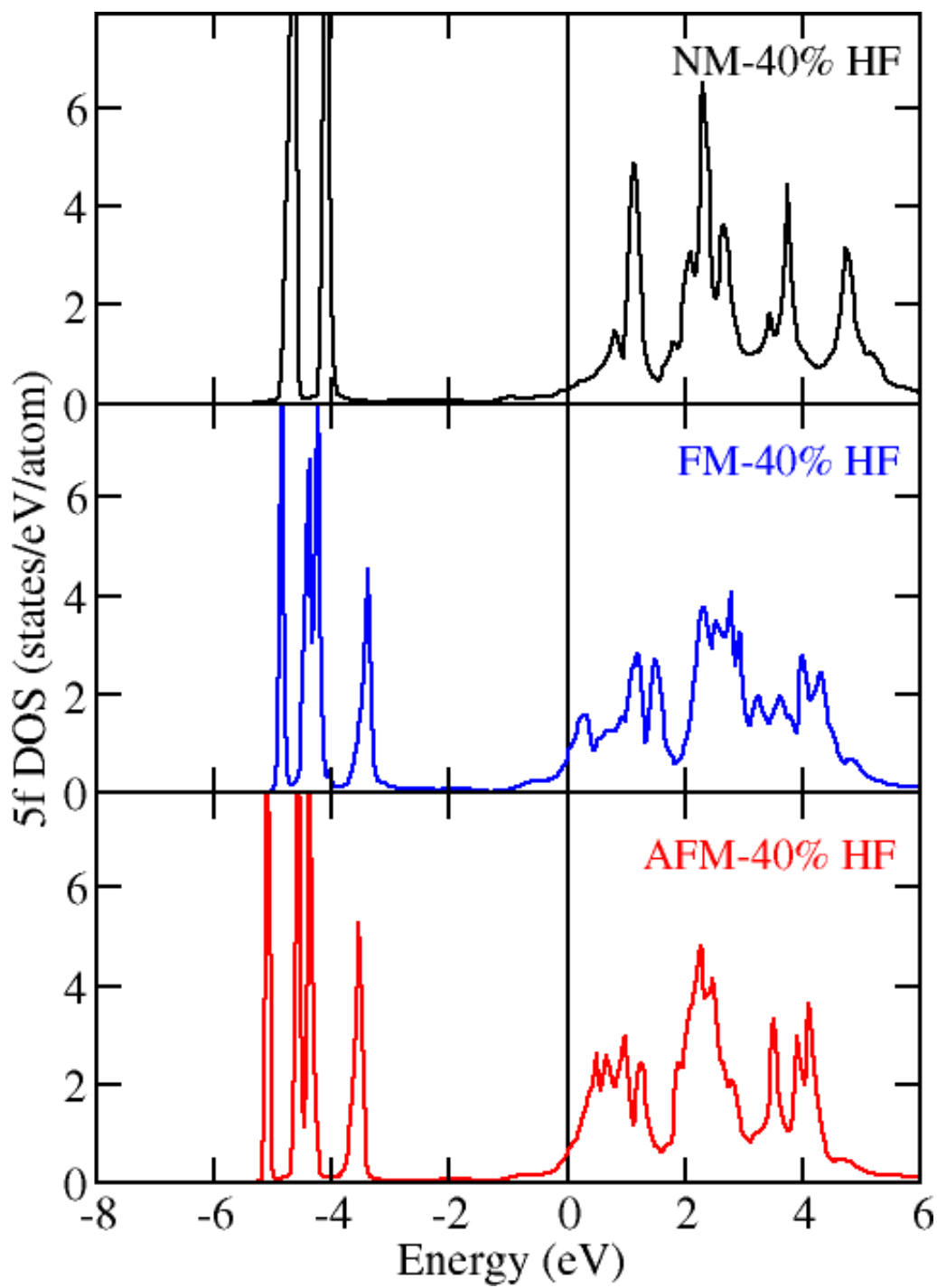

Fig. 4: 5*f* partial density of states for δ-Pu for the non-magnetic (NM), ferromagnetic (FM), and anti-ferromagnetic (AFM) computed using the PBE2 functional.

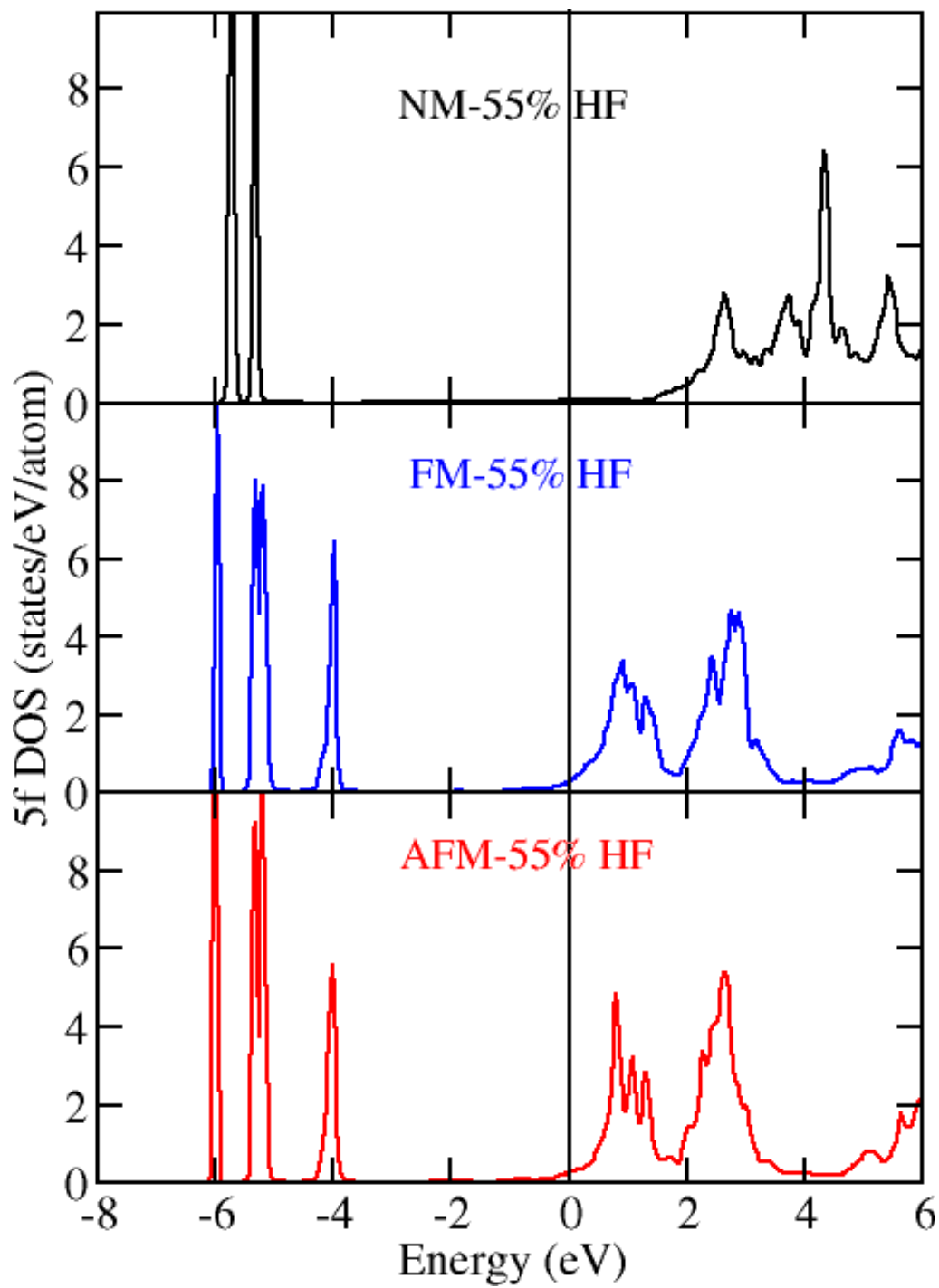

Fig. 5: *5f* partial density of states for δ-Pu for the non-magnetic (NM), ferromagnetic (FM), and anti-ferromagnetic (AFM) computed using the PBE3 functional.